\documentclass[onecolumn, manuscript]{revtex4}
\usepackage{graphicx}
\begin{document}

\title{Energy diffusion in strongly driven quantum chaotic systems}
\author{P.V. Elyutin}
\email{pve@shg.phys.msu.su} \affiliation {Department of Physics,
Moscow State University, Moscow 119992, Russia}
\date{\today}

\begin{abstract}
The energy evolution of a quantum chaotic system under the
perturbation that harmonically depends on time is studied for the
case of large perturbation, in which the rate of transition
calculated from the Fermi golden rule exceeds the frequency of
perturbation.  It is shown that the energy evolution retains its
diffusive character, with the diffusion coefficient that is
asymptotically proportional to the magnitude of perturbation and
to the square root of the density of states.  The results are
supported by numerical calculation.  They imply the absence of the
quantum-classical correspondence for the energy diffusion and the
energy absorption in the classical limit $\hbar \to 0$.

\vspace{10mm} PACS numbers {05.45.-a}
\end{abstract}
\maketitle

\section{Introduction}
The problem of susceptibility of chaotic  systems to perturbations
has attracted much attention in the last decade [1 - 9].  This
problem is fundamental, since it includes the determination of the
response of a material system to an imposed external
electromagnetic field, the setup that is typical for many
experiments.  Due to the sensitivity of classical phase
trajectories or quantum energy spectra and stationary
wavefunctions of chaotic systems to small changes of their
parameters, the problem is challengingly difficult.  A consistent
and noncontroversial picture covering (albeit qualitatively) all
the essential cases of the response hasn't been yet drawn at
present. From the point of view of general theory, the problem is
related to the applicability of the concept of quantum-classical
correspondence to chaotic systems, that is a long-standing
question in its own right \cite{E88,+E88}.

We shall study a one-particle system with the Hamiltonian of the
form $\hat H = \hat H_0  - F\hat x\cos \omega _0 t$, where $\hat
H_0 \left( {{\bf{\hat p}},{\bf{\hat r}}} \right)$ is the
Hamiltonian of the unperturbed system; $\bf{\hat p}$  and
$\bf{\hat r}$ are the operators of Cartesian components of the
momentum and of the position of the particle.  The classical
system with the Hamiltonian function $H_0 \left(
{{\bf{p}},{\bf{r}}} \right)$ will be assumed to be strongly
chaotic, that is, nearly ergodic on the energy surfaces in a wide
range of the energy values, system with $d \ge 2$ degrees of
freedom.  In the perturbation operator $\hat V\left( t \right) = -
F\hat x\cos \omega _0 t$ the active variable $\hat x$ is one of
the Cartesian coordinates of the particle, coupled to the external
homogeneous force field.  The amplitude $F$ in the following will
be referred to as field.  In the following we shall deal with the
quasiclassical case, when the Planck constant  is small in
comparison of the action scale of the system $H_0$.

Under the influence of the perturbation the energy value
$E(t)=H_0(t)$ varies in a quasirandom way.  These variations
frequently can be described as a process of the energy diffusion
\cite{LG91,J93}, when for the ensemble with the microcanonical
initial energy distribution $H_0(0)=E$ the dispersion of the
energy increases linearly with time, $\left\langle {\Delta E^2
\left( t \right)} \right\rangle  = 2Dt$, where $D\left(
{E,F,\omega _0 } \right)$ is the energy diffusion coefficient.

If the external field $F$ is sufficiently small in comparison with
the appropriately averaged values of the forces acting on a
particle in the unperturbed system, then in the classical model
the energy diffusion coefficient $D$ can be expressed through the
characteristics of the unperturbed chaotic motion of the active
coordinate, namely
\begin{equation}\label{1}
D = \frac{\pi }{2}\omega _0^2 F^2 S_x \left( {E,\omega _0 }
\right),
\end{equation}
where $S_x(E,\omega_0)$ is the power spectrum of the active
coordinate (the Fourier transform of its autocorrelation function)
for the motion on the surface with the constant energy value $E$
[9].    The same expression (1) in the case of weak perturbation
can be obtained in the classical limit from the quantum model. The
evolution of the quantum system can be treated as a sequence of
one-photon transitions between stationary states of the
unperturbed system $\left| n \right\rangle  \to \left| k
\right\rangle $, accompanied with the absorption or emission of
the quanta $\hbar \omega _0 $. For small $\hbar$ the energy
spectrum of $\hat H_0 $ is quasicontinuous, thence the rates of
transition are given by the Fermi golden rule (FGR)
\begin{equation}\label{2}
\dot W_F  = \frac{\pi }{{2\hbar }}F^2 \left| {x_{nk} } \right|^2
\rho \left( {E_k } \right),
\end{equation}
where $x_{nk}$ is the matrix element of the active coordinate, and
$\rho(E_k)$ is the density of states near the final state of the
transition.  Although the matrix elements $x_{nk}$  in quantum
chaotic systems fluctuate wildly with the variation of $k$ [10,
11], the averaged squared quantity $\overline {\left| {x_{nk} }
\right|^2 } $ in the limit $\hbar  \to 0$ is smooth; it is
proportional to the power spectrum $S_x \left( {E,\omega _0 }
\right)$ of the coordinate \cite{FP86,W87},
\begin{equation}\label{3}
\overline {\left| {x_{nk} } \right|^2 }  \approx \frac{{S_x \left(
{E,\omega _0 } \right)}}{{\hbar \rho \left( E \right)}}.
\end{equation}
From Eqs. (2) and (3) we have for the transition rate
\begin{equation}\label{4}
\dot W_F  = \frac{\pi }{{2\hbar ^2 }}F^2 S_x \left( {E,\omega }
\right).
\end{equation}
Then for the energy dispersion for small $t$ we have $\left\langle
{\Delta E^2 } \right\rangle  = 2\left( {\hbar \omega _0 }
\right)^2 \dot W_F t$, that returns us to the Eq.(1) for the
energy diffusion coefficient.  It can be shown that the same
expression for $D$ holds also for large $t$ \cite{+E04}. The
energy absorption in chaotic systems comes as an epiphenomenon of
the energy diffusion [4].  With the account of the dependence on
the energy of the power spectrum $S_x \left( {E,\omega } \right)$
and the density of states $\rho \left( E \right)$ the diffusion
becomes biased, and the energy absorption rate $Q$ is given by the
formula \cite{+ESh96, C99}
\begin{equation}\label{5}
Q = \frac{1}{\rho }\frac{d}{{dE}}\left( {\rho D} \right).
\end{equation}

Although for weak fields $D$ does not depend on the Planck
constant $\hbar$, the condition of the applicability of Eq. (2)
does.  The FGR is, after all, only a formula of the first order
perturbation theory.  It is based on the assumption that the
transition process has a resonant character - that the width
$\Delta$ of the energy distribution of states populated from the
original one, given by the Weisskopf - Wigner formula $\Delta  =
\hbar \dot W$ \cite{WW30}, is small in comparison with the quanta
energy $\hbar \omega _0 $.  From Eq. (4) it is evident that in the
classical limit $\hbar  \to 0$ this condition will be violated. In
the following we shall use the border value of the field $F_b$,
defined by the condition $\dot W_F \left( {F_b } \right) = \omega
_0 $, and refer to the domain $F \ge F_b $ as the range of the
strong field.

By analogy with other models, for strong fields one can expect a
slow-down of the growth of the energy diffusion coefficient $D$
and of the energy absorption rate $Q$.  For example, for a
two-level system with relaxation the quadratic dependence the
absorption rate $Q \propto F^2 $ for small field turns into a
field-independent value $Q_0$ for strong one.  The border is
determined by the condition  ${{\Omega ^2 } \mathord{\left/
{\vphantom {{\Omega ^2 } {\Gamma _1 \Gamma _2 }}} \right.
\kern-\nulldelimiterspace} {\Gamma _1 \Gamma _2 }}\sim 1$, where
$\Omega$ is the Rabi frequency and ${\Gamma _1 \Gamma _2 }$ are
longitudinal and transversal relaxation rates correspondingly
\cite{AE75}.  The rate of transitions from discrete to continuous
energy spectrum (that are basically covariant with the energy
absorption rate $Q$), studied in the context of the theory of
photoionization, for sufficiently strong fields can even decrease
with the increase of $F$ - the effect that is known as atom
stabilization by the strong field \cite{DK95}.

For our model we can describe the slow-down by a simple estimate.
Assuming the exponential decay of the initially populated state
and using the Weisskopf - Wigner formula for the energy
distribution of the final states [16], we can rewrite the
expression for the rate of the transitions (with the absorption of
quanta) in the form
\begin{equation}\label{6}
\dot W = \frac{\pi }{2}\int {\frac{{V^2 \left( \omega
\right)}}{\hbar }\left[ {\frac{1}{\pi } \cdot \frac{{\dot
W}}{{\left( {\omega  - \omega _0 } \right)^2  + \dot W^2 }}}
\right]} \,\rho \left( \omega  \right)d\omega,
\end{equation}
where $V(\omega)$ is a matrix element of perturbation taken as a
function of the transition frequency and $\omega  = {{\left( {E_k
- E_n } \right)} \mathord{\left/  {\vphantom {{\left( {E_k  - E_n
} \right)} \hbar }} \right. \kern-\nulldelimiterspace} \hbar }$.
In the weak field the expression in square brackets turns into the
$\delta$ - function, and Eq. (6) returns the usual FGR.  However,
if we dare to use Eq. (6) as an equation that is valid for any
magnitude of the perturbation, then in the case of strong fields
we will have
\begin{equation}\label{7}
\dot W^2  = \frac{1}{2}\int {\frac{{V^2 \left( \omega
\right)}}{\hbar }} \,\rho \left( \omega  \right)d\omega.
\end{equation}
The rate of transitions $\dot W$, and, consequently, the energy
diffusion coefficient $D$ and the energy absorption rate $Q$,
become proportional to the magnitude of perturbation $F$ and to
the square root of the density of states.  This derivation of Eq.
(7) is too risky - but in the following sections this conclusion
will be put on a more solid footing.

The slow-down of the energy diffusion in quantum chaotic systems
in strong harmonic fields has been first demonstrated by Cohen and
Kottos \cite{CK00}.  However, their analytical estimates and data
of numerical experiments are in quantitative disagreement with the
results of the present paper.

Lastly, it must be noted that the strong field regime is easily
attainable to experiments.  For example, for the excitation of
multiatomic molecules with the infrared laser radiation the border
field corresponds to the intensity value  $I \sim 10^9
\,{\rm{W}}\,{\rm{cm}}^{{\rm{ - 2}}} $ , that has been reached in
experiments long time ago \cite{LM81}.

\section{Theory}

For the system with the Hamiltonian $\hat H = \hat H_0  + \hat
V\cos \omega _0 t$ we take the wave function in the form of the
expansion in the basis of stationary $\left\{ {\varphi _m }
\right\}$ states of  $\hat H_0 $,
\begin{equation}\label{8}
\Psi \left( t \right) = \sum\limits_m {a_m \varphi _m \left(
{\bf{r}} \right)\,} e^{ - i\omega _m t} .
\end{equation}
Then for the amplitudes $\left\{ {a_m \left( t \right)} \right\}$
we obtain the system of equations
\begin{equation}\label{7}
i\frac{{da_k }}{{dt}} = \sum\limits_k {\Omega _{km} \cos \omega _0
t\,e^{i\omega _{km} t} a_m } ,
\end{equation}
where the quantities $\Omega _{kn}  = \hbar ^{ - 1} Fx_{kn}$ are
the Rabi frequencies of transitions.  We shall use the initial
conditions $a_m \left( 0 \right) = \delta _{mn} $: at the initial
moment only one of stationary states, $\varphi_n$, is populated.
Following \cite{FP86, W87}, we shall assume that $x_{nk}$ are
independent random Gaussian variables with zero mean value and the
dispersion, given by Eq. (3).  The system of equations (9) will be
treated as a member of the corresponding statistical ensemble.

We are going to concentrate on the process of the energy
diffusion.  Then in the zeroth approximation we can restrict
ourselves by consideration of the evolution of probability density
in a narrow energy range around the initial state and use the
power spectrum and the density of states values at this energy:
$S_x \left( \omega  \right) \equiv S_x \left( {E_n ,\omega }
\right)$ and $\rho  \equiv \rho \left( {E_n } \right)$.  For the
calculation of the absorption coefficient the global dependence on
the energy must be restituted.

The power spectrum $S_x \left( \omega  \right)$ has the symmetry
property $S_x \left( { - \omega } \right) = S_x \left( \omega
\right)$.  The dependence $S_x \left( \omega  \right)$ in the
domain $\omega  > 0$ in typical strongly chaotic systems, such as
nonlinear oscillators [20] and billiards [21, 22], has the form of
an asymmetric peak.  We shall define the peak value of the Rabi
frequency simply as $\Omega$, the frequency of the maximum as
$\tilde \omega $ and the characteristic width of the peak as
$\Gamma$.  Typically the ratio ${{\tilde \omega } \mathord{\left/
{\vphantom {{\tilde \omega } \Gamma }} \right.
 \kern-\nulldelimiterspace} \Gamma }$ is about few units.

Right after the switching on the perturbation all amplitudes (but
that of the initially populated state) grows in absolute value
linearly in time.  At this ballistic stage the energy dispersion
grows quadratically in time:
\begin{equation}\label{10}
\left\langle {\Delta E^2 } \right\rangle  \approx K_1 \hbar ^3
\tilde \omega ^2 \Omega ^2 \Gamma \rho t^2 ,
\end{equation}
where $K_1$ is a numerical constant.  This stage is limited by the
depletion of the initial population and lasts until the depletion
time $t_d \sim \Omega ^{ - 1} \left( {\hbar \Gamma \rho }
\right)^{{{ - 1} \mathord{\left/  {\vphantom {{ - 1} 2}} \right.
\kern-\nulldelimiterspace} 2}}$.  At this time considerably
populated levels are spread over the energy range $\Delta E\sim
\hbar \tilde \omega $ that contains many levels (since $\rho
\propto \,\hbar ^{ - d} $ with $d \ge 2$).  We shall expect that
at the next stage the ensemble averaged density of probability is
a smooth function with a characteristic scale $\Delta E >  > \hbar
\tilde \omega $.

It is convenient to write indices in Eq. (9) as arguments of
functions.  We shall use the frequency distance from the initial
level as a basic independent variable $\omega$, $a_k$ will be
denoted as $a(\epsilon)$, where $\varepsilon  = {{\left( {E_k  -
E_n } \right)} \mathord{\left/ {\vphantom {{\left( {E_k  - E_n }
\right)} \hbar }} \right.  \kern-\nulldelimiterspace} \hbar }$.
Dummy variables $\eta$ and $\eta '$ will have the same meaning.
By formal integration of Eq. (9) and subsequent recurrent
substitution we obtain the equation for the rate of change of the
local probability density $w\left( \varepsilon  \right) = \left|
{a\left( \varepsilon  \right)} \right|^2 $:

\begin{eqnarray}\label{11}
\frac{{dw\left( \varepsilon  \right)}}{{dt}} = \sum\limits_{\eta ,\eta '}
{\Omega \left( {\varepsilon ,\eta } \right)} \,e^{i\left( {\varepsilon  -
\eta } \right)t} \cos \omega _0 t\,a\left( {\eta ,t} \right) \times  \\
\,\,\,\,\,\,\,\,\,\,\,\,\,\,\,\,\,\,\,\,\,\,\,\,\, \times
\int\limits_{}^t {dt'\Omega \left( {\varepsilon ,\eta '}
\right)\,} e^{ - i\left( {\varepsilon - \eta '} \right)t} \cos
\omega _0 t\,a^ *  \left( {\eta ',t} \right) +
{\rm{c}}{\rm{.c}}{\rm{.}}\nonumber
\end{eqnarray}
Summation in this formula goes over discrete values of $\eta$ and
$\eta '$, and this equation is still exact.

Now we assume, that the amplitudes $a\left( {\eta ,t} \right)$ are
random processes, that are not correlated for different states:
$\left\langle {a\left( {\eta ,t} \right)a^ *  \left( {\eta ',t'}
\right)} \right\rangle  \propto \delta _{\eta \eta '} $.  Then for
the averaged probability density we can retain in the Eq. (11)
only diagonal terms:
\begin{eqnarray}\label{12}
\frac{{d\left\langle {w\left( \varepsilon  \right)} \right\rangle
}} {{dt}} = \sum\limits_\eta  {\left\langle {\Omega ^2 \left(
{\varepsilon ,
\eta } \right)} \right\rangle } \,\cos \omega _0 t \times  \\
\times \int\limits_{}^t {dt'\,} e^{i\left( {\varepsilon  - \eta }
\right)\left( {t - t'} \right)} \cos \omega _0 t'\,\left\langle
{a\left( {\eta ,t} \right)a^ *  \left( {\eta ,t'} \right)}
\right\rangle  + {\rm{c}}{\rm{.c}}{\rm{.}} \nonumber
 \end{eqnarray}

With the assumption that the averaged $\left\langle {w\left(
{\varepsilon ,t} \right)} \right\rangle $ is a smooth function of
$\varepsilon $, and slowly varying function of $t$, we may cast
the product of amplitudes in the form
\begin{equation}\label{13}
\left\langle {a\left( {\eta ,t} \right)a^ *  \left( {\eta ,t'}
\right)} \right\rangle  + \textrm{c.c.} = 2\left\langle {w\left(
{\eta ,t} \right)} \right\rangle B\left( {t - t'} \right),
\end{equation}
where $B\left( \tau  \right)$ is the normalized ($B\left( 0
\right) = 1$) autocorrelation function of the probability
amplitudes.

By replacing the averaged square Rabi frequency by its value from
Eq. (3) (that depends only on the difference $\varepsilon  - \eta
$), substituting the summation over the states by the integration
weighted with the density of states, and averaging over the time
intervals that are much larger than the field period,  we come to
the equation
\begin{equation}\label{14}
\frac{{dw\left( \varepsilon  \right)}}{{dt}} = \int {d\eta \Omega
^2 } \left( {\varepsilon  - \eta } \right)\rho \left( \eta
\right)\int\limits_0^\infty  {d\tau \cos \left( {\varepsilon  -
\eta } \right)\tau \cos \omega _0 \tau B\left( \tau  \right)}
w\left( \eta  \right).
\end{equation}
From now we drop the angular brackets and deal only with ensemble
averaged quantities.  If the rate of variations of $w\left( {\eta
,t} \right)$ is small in comparison with the decay of correlations
of amplitudes given by $B\left( \tau  \right)$, then we can treat
Eq. (13) as summation over the probability flow that comes from
the different parts of the frequency range with the constant rate,
\begin{equation}\label{15}
\dot W\left( {\eta  \to \varepsilon } \right) = \Omega ^2 \left(
{\varepsilon  - \eta } \right)\int\limits_0^\infty  {d\tau \cos
\left( {\varepsilon  - \eta } \right)} \tau \cos \omega _0 \tau
B\left( \tau  \right).
\end{equation}
This approximate expression to some extent replaces the Fermi
golden rule for strong perturbations.

To construct the kinetic equation one must take into account both
incoming and outgoing flows of probability.  By taking into
account the total probability flow, expanding $w\left( \varepsilon
\right)$ in the Taylor series, we obtain a diffusion equation with
the probability diffusion coefficient in the energy scale
\begin{equation}\label{16}
D \approx \int\limits_{ - \infty }^\infty  {d\eta \hbar ^3 \eta ^2
\Omega ^2 \left( \eta  \right)\rho \,J\left( {\eta ,\omega _0 }
\right)},
\end{equation}
where
\begin{equation}\label{17}
J\left( {\eta ,\omega _0 } \right) = \int\limits_0^\infty  {d\tau
\cos \eta } \tau \cos \omega _0 \tau B\left( \tau  \right).
\end{equation}

Now the problem reduces to the calculation of the integral
$J\left( {\eta ,\omega _0 } \right)$.  For large enough times the
average probability density, that is governed by the diffusion
equation, varies slowly, and we can treat the system Eq. (9) as a
set of equations in which all $a_m \left( t \right)$ are
non-correlated random processes with the same statistical
properties.  Then by averaging the equation for the squared time
derivative of an amplitude, we obtain the expression for the mean
squared frequency of these processes, that also gives the estimate
for the square of the correlation decay rate $\gamma$:
\begin{equation}\label{18}
\left\langle {\omega ^2 } \right\rangle  = \frac{1}{2}\int {\Omega
^2 \left( \eta  \right)\hbar \rho d\eta }  \equiv \gamma ^2.
\end{equation}

From Eq. (18) for the decay correlation rate we have the estimate
\begin{equation}\label{19}
\gamma  \approx K_2 \Omega \sqrt {\Gamma \hbar \rho }\, ,
\end{equation}
where $K_2$ is a numerical constant  In the domain of strong field
the autocorrelation function is the fastest component under the
integration sign in Eq. (17).  Then we have $J\sim \gamma ^{ - 1}
\approx \left( {K_2 \Omega \sqrt {\mathstrut \Gamma \hbar \rho } }
\right)^{ - 1} $.  By the substitution of this expression into Eq.
(16) we obtain the estimate of the energy diffusion coefficient:
\begin{equation}\label{20}
D \approx K_3 \hbar ^2 \tilde \omega ^2 \Omega \sqrt {\Gamma \hbar
\rho },
\end{equation}
where $K_3$ is a numerical constant; this expression agrees with
the dependence that was derived in Sec. 1.  One must recall that
this expression is valid only for the nearly resonant perturbation
frequency  ( $\left| {\omega _0  - \tilde \omega } \right| \le
\Gamma $ ).  We do not enter here into the studies of the
dependence of $D$ on the perturbation frequency $\omega_0$ in the
wider domain, postponing it for the further studies.

\section{Numerical experiment}

To check the analytical results of the preceding section, the
system of equations (9) has been integrated numerically.  The
number of equations varied from $N=300$ to $N=1200$ with the
purpose to suppress the influence of the border.  The envelope of
the Rabi frequencies has been taken in the double lorentzian form
\begin{equation}\label{21}
\Omega _{m,n}  = \Omega \left[ {\frac{{\Gamma ^2 }}{{\left(
{\omega _m  - \omega _n  + \tilde \omega } \right)^2  + \Gamma ^2
}} + \frac{{\Gamma ^2 }}{{\left( {\omega _m  - \omega _n  + \tilde
\omega } \right)^2  + \Gamma ^2 }}} \right].
\end{equation}
All calculations have been carried for the "resonant" perturbation
frequency $\omega _0  = \tilde \omega $ and for the peak width
$\Gamma  = 0.3\tilde \omega $.

\begin{figure}
[!ht]
\includegraphics[width=0.8\columnwidth]{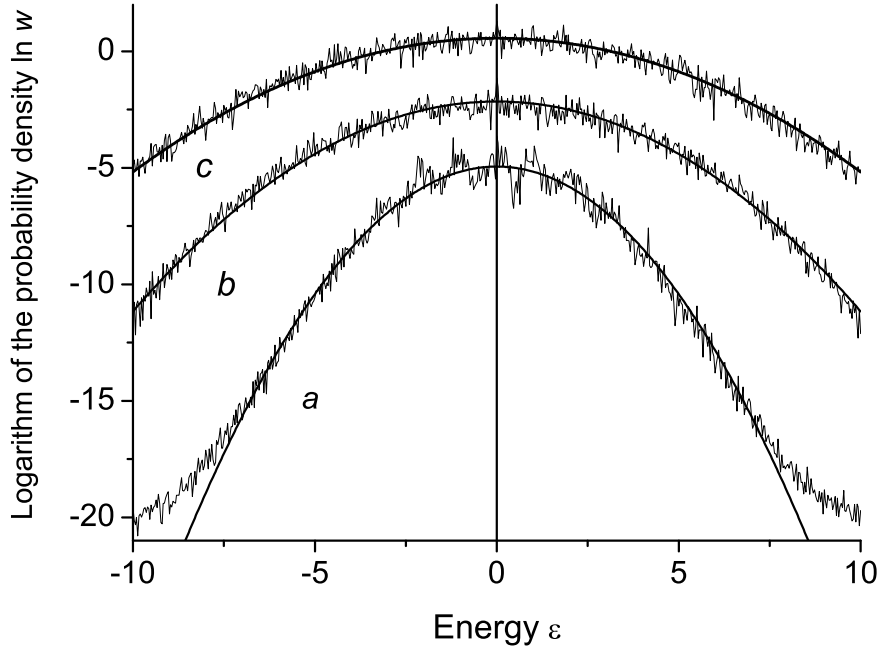}
\caption{\label{fig1} The dependence of the logarithm of the
probability density on the dimensionless frequency $\varepsilon '
=  {\varepsilon} / {\tilde \omega}$ for the time values $t =
5\tilde \omega ^{ - 1} $ (a), $t = 10\tilde \omega ^{ - 1} $ (b),
and  $t = 15\tilde \omega ^{ - 1} $ (c).  The grassy lines -
values of $\ln w\left( {\varepsilon '} \right)$, averaged over 10
different sets of matrix elements, solid lines - fitted parabolas.
To avoid the overlap of graphs, the plots for cases (b) and (c)
are shifted upwards by 3 and 6 units respectively.}
\end{figure}

Fig. 1 shows the distribution of probability as a function of
dimensionless frequency $\varepsilon ' = {\varepsilon
\mathord{\left/  {\vphantom {\varepsilon  {\tilde \omega }}}
\right. \kern-\nulldelimiterspace} {\tilde \omega }}$ for
different moments of time.  It is clearly seen that even for
relatively small time $t = 5\tilde \omega ^{ - 1}  = 2.2t_d $ the
distribution has very accurate Gaussian form, with deviations
noticeable only for $\left| {\varepsilon '} \right| \ge 7.5$. Thus
we have quantitative support for our conclusion about the
diffusive character of the energy evolution.

\begin{figure}
[!ht]
\includegraphics[width=0.8\columnwidth]{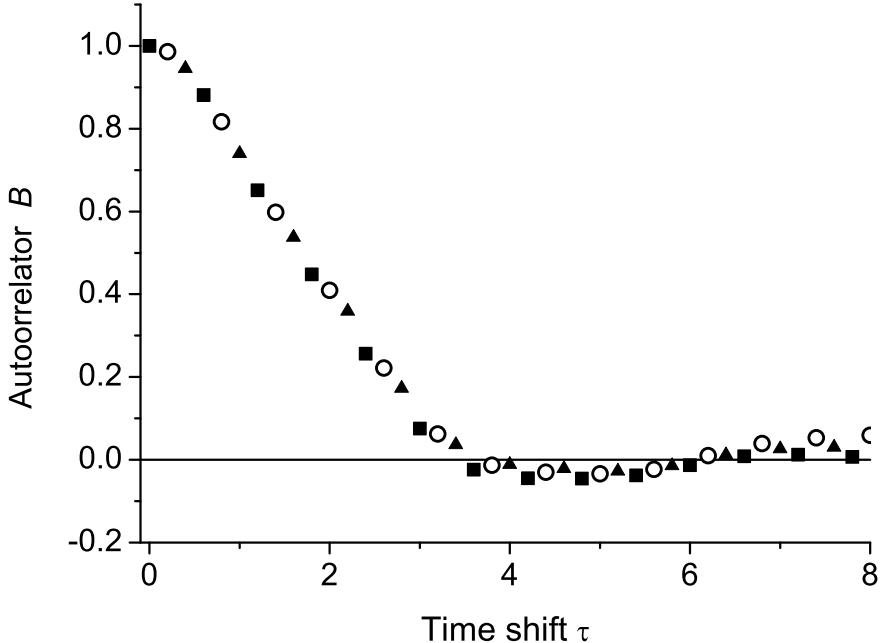}
\caption{\label{fig1} Dependence of the autocorrelation function
of the probability amplitudes on the dimensionless time shift
$\tau ' = \tau \tilde \omega $ for three different sets of
parameters with $\tilde \omega \hbar \rho  = 30$ (black squares),
$\tilde \omega \hbar \rho  = 60$ (open circles) and $\tilde \omega
\hbar \rho  = 120$ (black triangles) and the same value of $\Omega
\sqrt {\Gamma \hbar \rho }  = 0.618\tilde \omega $.  The
statistical errors are about the size of the data symbols. }
\end{figure}

Fig 2 depicts the form of the normalized autocorrelation function
of the probability amplitudes.  Values $B\left( \tau  \right)$
have been calculated numerically for three sets of parameters with
different values of $\rho$ but the same value of the product
$\Omega \sqrt {\hbar \Gamma \rho } $.  It is clearly seen that
$B\left( \tau  \right)$ for these sets are nearly identical, as
expected.  The decay rate $\gamma$ taken from the equation
$B\left( {\gamma ^{ - 1} } \right) = \exp \left( { - 1} \right)$
is $\gamma  = 0.77\,\Omega \sqrt {\mathstrut \hbar \Gamma \rho }\,
$, that supports the estimate Eq. (19).

\begin{figure}
[!ht]
\includegraphics[width=0.8\columnwidth]{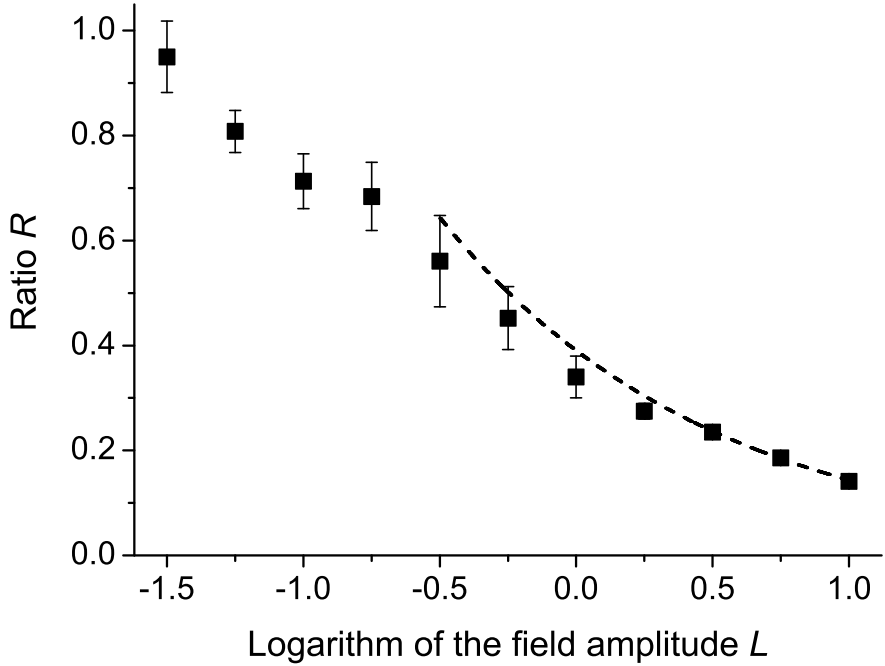}
\caption{\label{fig1} The dependence of the ratio $R = D / {2\dot
W_F} $ of the energy diffusion constant $D$ to the doubled Fermi
transition rate $\dot W_F  = (\pi/2)\Omega ^2 \hbar \rho $  on the
logarithm of the ratio of the Rabi frequency to its border value
$L = \Omega / \Omega_b$, $\Omega_b=(2 \tilde \omega / \pi \hbar
\rho)^{1/2}$. The dashed line represents the curve $R = A\exp
\left( { - L} \right)$ that corresponds to the theoretical
dependence Eq. (20); it is fitted to the last three points.}
\end{figure}

\newpage

Figure 3 represents the dependence of the ratio $R = {D
\mathord{\left/ {\vphantom {D {2\dot W_F }}} \right.
\kern-\nulldelimiterspace} {2\dot W_F }}$ of the energy diffusion
constant $D$ to the doubled Fermi transition rate $\dot W_F  =
\left( {{\pi  \mathord{\left/ {\vphantom {\pi  2}} \right.
\kern-\nulldelimiterspace} 2}} \right)\Omega ^2 \hbar \rho $ on
the logarithm of the ratio of the Rabi frequency to its border
value $L = {\Omega  \mathord{\left/ {\vphantom {\Omega  {\Omega _b
}}} \right. \kern-\nulldelimiterspace} {\Omega _b }}$, $\Omega _b
= \left( {{{2\tilde \omega } \mathord{\left/  {\vphantom {{2\tilde
\omega } {\pi \hbar \rho }}} \right.  \kern-\nulldelimiterspace}
{\pi \hbar \rho }}} \right)^{{1 \mathord{\left/ {\vphantom {1 2}}
\right. \kern-\nulldelimiterspace} 2}} $.  It is seen that for the
weak field this ratio comes close to the asymptotic limit (unity),
decreases in the vicinity of the border and decays as $F^{-1}$ for
sufficiently strong fields.  The agreement with the theoretical
estimates is quite convincing.

\section{Conclusion}

From the comparison of the numerical data with the theoretical
estimates we can conclude that the approach of the Sec. 2 gives
reasonably accurate description of the energy evolution process in
strong fields, in spite of numerous simplifying approximations
used in the calculation.

To improve the accuracy and, firstly, to derive from the first
principles the equation for the correlation function $B(\tau )$,
the model of the random process must be improved.  We used the
model of the stationary process, whereas from Eq. (9) one can
conclude that the model of periodic random process would be more
appropriate.

The main conclusion from the results of our calculation is a
qualitative one: by substituting Eq. (3) in Eq. (16), we obtain $D
\approx K_3 \hbar \tilde \omega ^2 F\left[ {S_x \left( {\omega _0
} \right)\Gamma } \right]^{{1 \mathord{\left/ {\vphantom {1 2}}
\right. \kern-\nulldelimiterspace} 2}} $.  This quantity in the
classical limit $\hbar  \to 0$ vanishes along with the energy
absorption rate $Q$ (see Eq. (5)).  That means the violation of
the quantum-classical correspondence for the absorption - and,
more generally, for the linear susceptibility of chaotic systems
to harmonic external fields.

\section*{Acknowledgements}
The author acknowledges the support by the "Russian Scientific
Schools" program (grant \# NSh - 1909.2003.2).

\end{document}